\documentclass{PoS}
\usepackage{slashed}

\title{Search for scalar bottom quarks and third-generation leptoquarks \\ in $p\overline{p}$ collisions at $\sqrt{s}$ = 1.96 TeV}

\ShortTitle{Search for scalar bottom quarks and third-generation leptoquarks at D0}

\author{C\'ecile Deterre, on behalf of the D0 collaboration\\
        CEA-Saclay (DSM/IRFU/SPP), France \\
        E-mail: \email{cecile.deterre@cea.fr}}

\abstract{We present the results of a search for pair production of scalar bottom quarks ($\tilde b_1$) and scalar
third-generation leptoquarks ($LQ_3$) in a data sample of 5.2 $fb^{-1}$ collected by the D0 experiment at the Tevatron, the $p\overline{p}$ collider at Fermilab. 
We assume that sbottoms decay to a neutralino ($\tilde \chi^0_1$) and a b quark,
and we set 95\% C.L. lower limits on their production in the ($m_{\tilde b_1}$, $m_{\tilde \chi^0_1}$) mass plane such that $m_{\tilde b_1} > 247$~GeV 
for $m_{\tilde \chi^0_1} = 0$ and $m_{\tilde \chi^0_1} > 110$~GeV for $160 < m_{\tilde b_1} < 200$~GeV. The leptoquarks are assumed
to decay to a tau neutrino and a b quark, and we set a 95\% C.L. lower limit of 247~GeV on the mass of a charge-1/3 third-generation leptoquark.
}

\FullConference{35th International Conference of High Energy Physics - ICHEP2010,\\
		July 22-28, 2010\\
		Paris France}

\begin{document}

In this search performed at the D0 experiment \cite{det}, we look for particles predicted by extensions of the Standard Model (SM) \cite{D0}: scalar bottom quarks in the framework of the Minimal Supersymetric Standard Model (MSSM) 
with R-parity conservation, and leptoquarks predicted by grand unified theories (GUT) and composite models.
In this analysis, we consider the region of parameter space 
where the only possible decay of the lighter sbottom quark $\tilde b_1$ is: $\tilde b_1 \to b \chi_1^0$.
$\chi_1^0$ is assumed to be the lightest supersymmetric particle and is therefore stable. 
Consequently, we look for: $p \overline{p} \to b\chi_1^0 \overline{b}\chi_1^0 $.
Charge-$\frac{1}{3}$ third-generation leptoquarks ($LQ_3$) are predicted to decay to $b\nu$ with the branching fraction B, and to $t\tau$ with the branching fraction 1-B.
We look for: $p \overline{p} \to LQ_3 \overline{LQ}_3 \to b\overline{b} \nu \overline{\nu}$.

\begin{figure}[h!]
 \centering
 \includegraphics[width=0.65\linewidth]{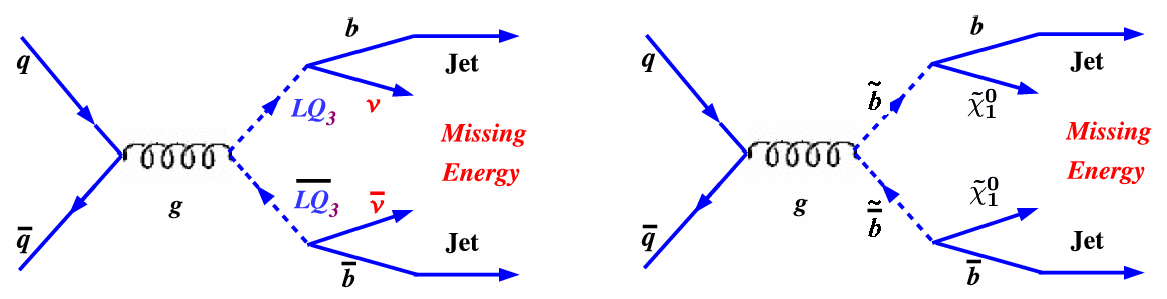}
 \caption{Feynman diagrams of the signals.}
 \label{fig:diag}
\end{figure}

The signal for both searches is two b-jets of high transverse momentum ($p_T$) and missing transverse energy ($\slashed{E}_T$) from escaping neutrinos or neutralinos (see figure~\ref{fig:diag}).

The main backgrounds come from processes with real $\slashed{E}_T$ such as decays of W and Z bosons with unidentified leptons, 
which are evaluated from Monte-Carlo simulations and normalized with a $W$ enriched sample of data.
Other backgrounds come from multijet processes with instrumental $\slashed{E}_T$ arising from energy mismeasurements and are evaluated from data with a QCD enriched sample.

We first select events with at least two jets each with $p_T$ > 20 GeV, and with an angle between the two leading jets of less than 165$^{\circ}$.
A veto is applied on events with isolated leptons to reduce the background from $W$ decays.
The background coming from multijet processes is reduced by applying a cut at 40~GeV on the $\slashed{E}_T$ and requiring it to be non-collinear with any jet. 
It is also required to have a high significance with respect to the resolution of the measured energies.
To further reduce that same background, a cut is applied on the angle between $\slashed{E}_T$ and the missing $p_T$ computed from the tracks, 
which tend to be aligned in events with real $\slashed{E}_T$.
At least two jets must be identified as b-jets, with one satisfying tight quality requirements.
Additional variables are used to reduce the contribution from events with poorly measured $\slashed{E}_T$, like the asymmetry 
$\mathcal{A}=\left(\slashed{E}_T - \slashed{H}_T \right) / \left(\slashed{E}_T + \slashed{H}_T \right)$ which has to be in the range $\left[-0.1,0.2 \right]$, where $ H_T = \sum_{jets} p_T$ and $ \slashed{H}_T = |\sum_{jets} \overrightarrow{p_T}|$.
The final selection consists of cuts on $\slashed{E}_T$, the leading jet $p_T$, $H_T$ and $X_{jj}~=~\left( p_T^{jet1} + p_T^{jet2} \right) / H_T$, 
which is used to reduce the background coming from top-quark processes.
The set of cuts depends on $m_{LQ_3}$ and $m_{\tilde b_1}$ and is optimized to yield the smallest expected limit on the cross section.

The main systematic uncertainties come from the uncertainty on the luminosity (6.1\%), the jet energy calibration and reconstruction (3\% for signal, 2 to 7\% for background), 
the b-tagging (6 to 17\% for signal, 5 to 11\% for background), the theoretical cross sections for the SM processes 
(10\% for the top background, and 6\% for W/Z bosons with an additional 20\% for heavy flavor content) and the contribution from the multijet background (25\%).

Using 5.2 fb$^{-1}$ of data, the number of events with the expected topology is consistent with the number of events predicted from SM processes (see Table~\ref{tab:yield}).

\begin{table}
 \centering
 \includegraphics[width=\textwidth]{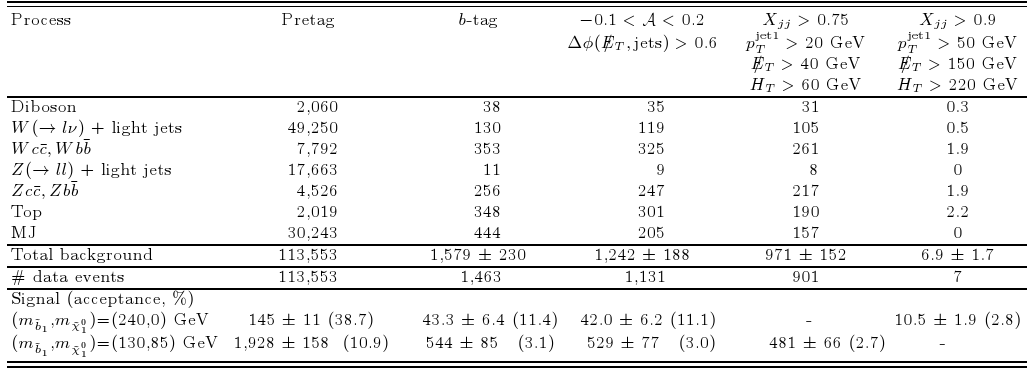}
 \caption{Predicted and observed number of events at different steps of the selection.}
 \label{tab:yield}
\end{table}

\begin{figure}[h]
 \begin{minipage}[c]{.48\linewidth}
   \includegraphics[width=\linewidth]{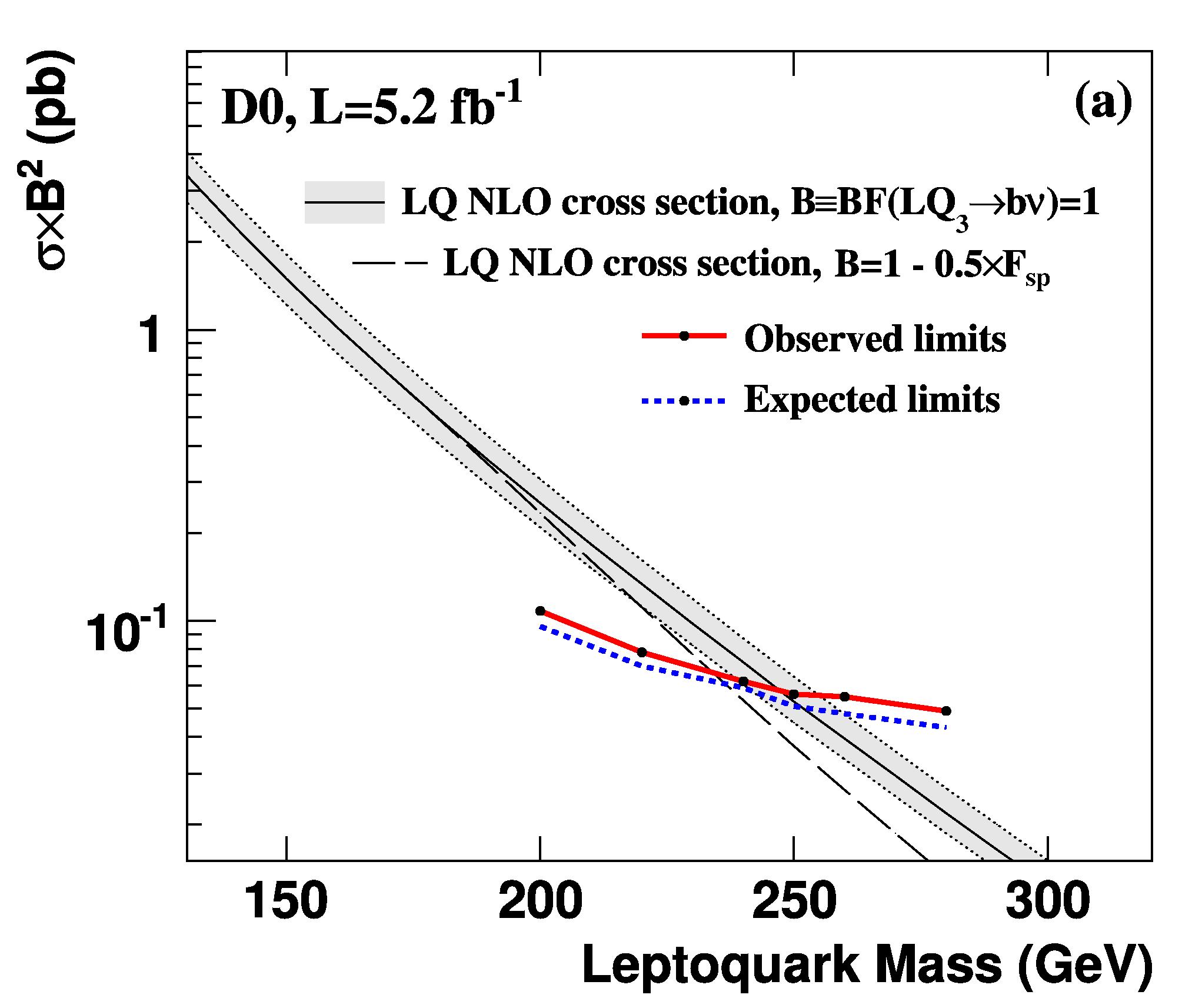}
 \end{minipage} \hfill
 \begin{minipage}[c]{.45\linewidth}
  \includegraphics[width=\linewidth]{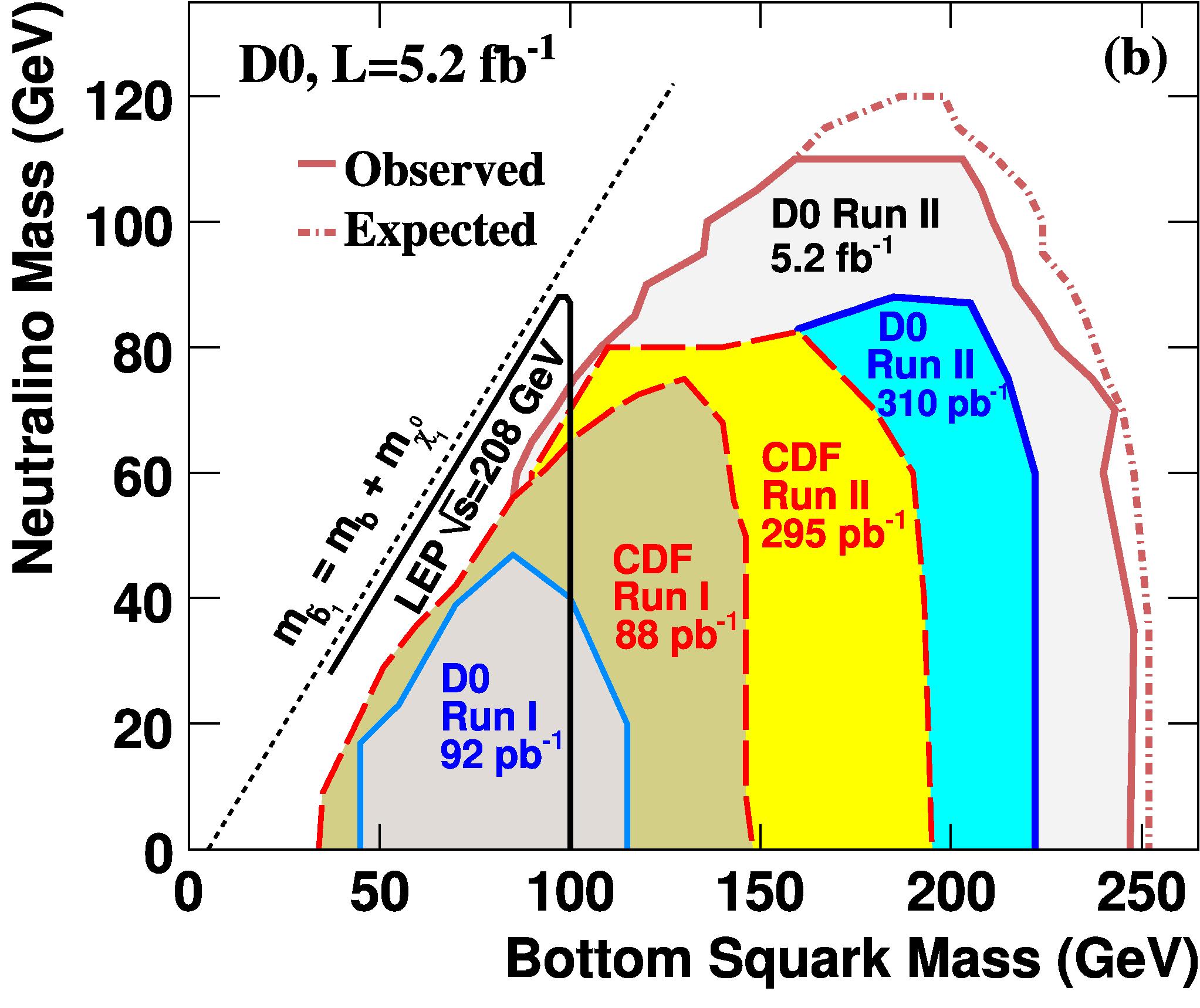}
 \end{minipage}
  \caption{(a) The 95\% C.L. limits on $\sigma B^2$ as a function of $m_{LQ_3}$ for the pair production of $LQ_3$.
           (b) The 95\% C.L. exclusion contour in the ($m_{\tilde b_1}$, $m_{\tilde \chi^0_1}$) plane. }
  \label{fig:conc}
\end{figure}

Figure \ref{fig:conc}(a) shows the 95\% C.L. upper limits on the cross section with respect to $m_{LQ_3}$ and the theoretical cross section assuming B = 1.
The limit on the mass obtained in this case for the production of leptoquarks of third-generation is $m_{LQ_3}$ > 247 GeV.
Also shown is the theoretical cross section when couplings to the $b\nu$ and $t\tau$ channels are identical.
In that case, the mass limit is 238 GeV.
Figure \ref{fig:conc}(b) shows the excluded region in the ($m_{\tilde b_1}$, $m_{\tilde \chi^0_1}$) plane.
For $m_{\tilde \chi^0_1} = 0$ the limit is $m_{\tilde b_1} > 247$~GeV at 95\% C.L.
For $160 < m_{\tilde b_1} < 200$~GeV, the limit is $m_{\tilde \chi^0_1} > 110$~GeV.
These limits significantly extend previous results.

\end{document}